\documentstyle[12pt,aasms4]{article}

\received{}
\accepted{}
\journalid{}{}
\articleid{}{}

\slugcomment{Accepted for publication in ApJ Main Journal}

\lefthead{B\'ejar, Zapatero Osorio, \& Rebolo}
\righthead{Very low-mass stars and brown dwarfs in the young 
$\sigma$\,Orionis cluster.}

\begin{document}

\title{A Search for Very Low-mass Stars and Brown Dwarfs in the Young 
$\sigma$\,Orionis cluster}

\author{V. J. S. B\'ejar, M. R. Zapatero Osorio, R. Rebolo}
\affil{Instituto de Astrof\'\i{}sica de Canarias, E-38200 La Laguna, 
Tenerife. Spain}

\centerline{e-mail addresses: vbejar@ll.iac.es, mosorio@ll.iac.es, 
rrl@ll.iac.es} 

\begin{abstract}
We present a CCD-based photometric survey covering 870\,arcmin$^2$ in a 
young stellar cluster around the young multiple star $\sigma$\,Orionis. 
Our survey limiting  $R$, $I$, and $Z$ magnitudes are 23.2, 21.8, and 21.0, 
respectively, with the completeness being about 2.2\,mag brighter. 
From our colour-magnitude diagrams, we have selected 49 faint objects 
($I$\,=\,15--21\,mag), which smoothly extrapolate the photometric sequence 
defined by more massive known members. Adopting the currently accepted 
age interval of 2--10\,Myr for the Orion 1b association, in which 
$\sigma$\,Orionis is located, and considering recent evolutionary models, 
our objects may span a mass range from 0.1 down to 0.02\,$M_{\odot}$, 
well within the substellar regime.  

Follow-up low-resolution optical spectroscopy (635--920\,nm) for eight of 
our candidates in the magnitude range $I$\,=\,16--19.5 shows that  they 
have spectral types M6--M8.5 which are consistent with the expectations 
for true members.  Compared with their Pleiades counterparts of similar 
types,  H$\alpha$ emission is generally stronger, while  Na\,{\sc i} and 
K\,{\sc i} absorption lines appear weaker, as expected for lower 
surface gravities and younger ages. Additionally, TiO bands and 
in particular VO bands appear clearly enhanced in our candidate with the 
latest spectral type, SOri\,45 (M8.5, $I$\,=\,19.5), compared to 
objects of similar types in older clusters and the field. We have estimated 
the mass of this candidate at only 0.020--0.040\,$M_{\odot}$, hence it is 
one of the least massive brown dwarfs yet discovered. 

We examine the potential role of deuterium as a tracer of both substellar 
nature and age in very young clusters. The luminosity and mass at which 
the burning/preservation of deuterium takes place is a sensitive function 
of age and  it can therefore provide a determination of the age of a 
cluster. The $\sigma$\,Orionis cluster is an excellent site for determining 
this transition zone empirically, where the most massive brown dwarfs 
identified  are expected to have burned their deuterium content, while 
the lowest mass ones should  have preserved it.    

\end{abstract}


\section{Introduction}

Stellar clusters and associations offer a unique opportunity to study 
substellar objects in a context  of known  age, distance and  metallicity; 
they are laboratories of key importance in  understanding the evolution of 
brown dwarfs. Deep imaging surveys have revealed a large population of 
substellar objects  in the Pleiades cluster (Rebolo,  Zapatero Osorio \& 
Mart\'{\i}n \cite{rebolo95};  Cossburn et al. \cite{cossburn97}; Zapatero 
Osorio, Rebolo \& Mart\'{\i}n \cite{osorio97a}; Zapatero Osorio et al. 
\cite{osorio97b}; Bouvier et al. \cite{bouvier98}), demonstrating that 
the formation of brown dwarfs extends down to masses of 0.035\,$M_{\odot}$ 
(Mart\'{\i}n et al. \cite{martin98a}; see also Luhman, Liebert \& Rieke 
\cite{luhman97}). The extension of these studies to other clusters, 
especially in the younger regions where we can reach objects with lower 
masses, is therefore  very important for confirming and enlarging these 
results. 

The Orion complex is recognized as one of the best sites for understanding 
star formation processes. Our knowledge of the young, low-mass stellar 
population in this star forming region has been enriched in recent years 
with the application of new search techniques for low-mass stars as, for 
example, H$\alpha$ surveys  (Wiramihardja et al. \cite{wiramihardja89}, 
\cite{wiramihardja91}, \cite{wiramihardja93}; Kogure et al. 
\cite{kogure89}), and the optical identification of X-ray sources, which 
were  detected by the {\it ROSAT} all-sky survey (RASS). These recent  
surveys provided a spatially unbiased sample of X-ray sources over the 
entire Orion complex (Sterzik et al. \cite{sterzik95}; Alcal\'a, 
Chavarr\'\i a-K., \& Terranegra \cite{alcala98}; Alcal\'a et al. 
\cite{alcala96}). {\it ROSAT} pointed observations performed on  
Orion's Belt (Walter et al. \cite{walter94}) led to the discovery of a 
high  concentration of X-ray sources near the bright O9.5V star 
$\sigma$\,Orionis. This star belongs to the Orion 1b association, for 
which an age of 1.7--7\,Myr and a distance modulus of 7.8--8 are 
estimated (Blaauw \cite{blaauw64};  Warren \& Hesser \cite{warren78}, 
hereafter WH; Brown, de Geus \& de Zeeuw \cite{brown94}, hereafter BGZ).  
Follow-up photometry and spectroscopy (Wolk \cite{wolk96}) of these X-ray 
sources have revealed a population of low-mass stars defining a 
photometric sequence consistent with the existence of a very young 
cluster at an age of few million years. This cluster is ideally suited 
for the detection of very low-mass brown dwarfs and subsequently for 
investigating the initial mass function in the substellar regime. 
Additionally, the multiple star $\sigma$\,Orionis is affected by an 
extinction of $E(B-V)$~=~0.05\,mag (Lee \cite{lee68}), and thus, the 
associated cluster may exhibit very little reddening.
At these early ages brown dwarfs are intrinsically more luminous (Burrows 
et al. \cite{burrows97}; D'Antona \& Mazzitelli \cite{dantona94}),  which 
makes their detection and study easier. For example, a 0.025\,$M_{\odot}$ 
object is about 7\,mag  brigh\-ter in the absolute magnitude M$_{I}$ at the 
age of 5\,Myr  than at that of the Pleiades cluster (120\,Myr) 
according to the recent tracks of Baraffe et al. (\cite{baraffe98}). In this 
paper we present the results of a deep photometric survey of the young 
cluster around the $\sigma$\,Orionis star in search of its substellar 
population. We also present follow-up low-resolution spectroscopy of 
some of our photometric candidates; and we discuss the possibility 
of using deuterium in studying very young substellar populations.  

\section{Observations}
\subsection{Photometry}
We have obtained $RIZ$ images with the Wide Field Camera (WFC), mounted at 
the prime focus of the Isaac Newton Telescope (INT), at Observatorio del 
Roque de los Muchachos (ORM) on the island of La Palma, on 1997  November 
29.  The camera consists of a mosaic of four 2048$\times$2048\,pixel$^2$ 
Loral CCD detectors, providing a pixel projection of 0.37\,arcsec. At the 
time of our observations, one of the CCDs was not in operation, so the 
effective area of one single mosaic was 478\,arcmin$^2$. We observed two 
different regions in each of the three filters,  covering a total area of 
870\,arcmin$^2$.  Figure~\ref{fig1} shows the location of the two mosaic 
fields (six CCDs) surveyed.  Exposure times were 1200\,s for all three 
filters for the east region and 2$\times$1200\,s for the west region.

Raw frames were reduced within the IRAF\footnote{IRAF is distributed by 
National Optical Astronomy Observatories, which is operated by the 
Association of Universities for Research in Astronomy, Inc., under 
contract with the National Science Foundation.} environment, using the 
CCDRED package. Images were bias-subtracted and flat-fielded. Due to poor 
weather conditions at dusk and dawn, it was not possible to take good 
sky flats.  We combined our long-exposure scientific images to obtain the 
flat-fields we finally used. The photometric analysis was performed using 
routines within DAOPHOT, which include the selection of stars using the  
DAOFIND routine (extended objects were mostly avoided)  and aperture and 
psf photometry. Observations at the INT  were affected  by cirrus; 
therefore no photometric standard stars were observed at this stage. 
Average seeing ranged from 1.3 to 2.0\,arcsec.  In order to transform 
our INT instrumental magnitudes into the Cousins $RI$ system, \- we made 
use of objects in common with images of our fields taken under photometric 
conditions with the IAC80 Telescope (Teide Observatory on the island of 
Tenerife) in  1998 January. The IAC80 images were calibrated with  
standards of Landolt (\cite{landolt92}).  We estimate the 1$\sigma$ 
error in our calibration to be around 0.1\,mag. Due to variability in 
weather conditions at the INT and to the different sensitivity among the 
CCDs of the WFC  our limiting magnitudes differ slightly from one image 
to the next. The survey  completeness magnitudes are $R$\,=\,20.5, 
$I$\,=\,19.5, and $Z$\,=\,19.2,  while limiting magnitudes are 
$R$\,=\,23.2, $I$\,=\,21.8, and $Z$\,=\,21.0. In Table~\ref{tab1} we 
list the limiting and completeness magnitudes for each of the CCDs and 
regions observed. 

For each CCD analyzed we constructed  the $I$ vs. $R - I$ and $I$ vs. 
$I-Z$ color--magnitude (CM) diagrams. These proved useful in 
distinguishing between cool cluster-member candidates and field  stars.  
Figure~\ref{fig2} shows the resulting $I$ vs. $R - I$ diagram of our 
survey where we have plotted  an arbitrary straight line separating 
our candidates from field  stars. Cluster-member candidates are 
identified as objects brighter and redder than the Pleiades sequence 
shifted to  the distance of $\sigma$\,Orionis in both CM diagrams. We 
have selected 46 very low-mass stars and brown dwarf candidates with 
magnitudes in the interval $I$\,=\,15--20\,mag that were found to be 
red in both  CM diagrams. Table~\ref{tab2} shows the list of our 
candidates with their magnitudes and coordinates (IAU designations are 
included). Those objects which do not appear red in both CM diagrams 
(they are indicated with different symbols in Fig.~\ref{fig2}) have not 
been considered in the following discussions on cluster membership and 
therefore, they are not listed in Table~\ref{tab2}. Nevertheless, it 
has been shown that the $R-I$ color saturates at a given value 
($R-I \sim 2.4$, Bessell \cite{bessell91}) becoming bluer for cooler 
objects. This value may be dependent on gravity, younger objects saturating 
at slightly redder colors (see Mart\'\i n, Rebolo, \& Zapatero Osorio 
\cite{martin96}; Bouvier et al. \cite{bouvier98}). Thus, $R-I$ cannot 
be used by itself as a good indicator of very faint cluster members, 
and we rely on the $I-Z$ color for the selection of candidates. The objects 
in our survey with $I$ magnitudes fainter than 20\,mag showing red $I-Z$ 
colors and $(R-I) \ge 2.1$ are listed in Table~\ref{tab3} (coordinates 
and IAU designations are given). This table may be not complete since 
the magnitude range here is clearly beyond the completeness 
of our survey. Astrometry was carried out using the USNO-SA2.0 catalog 
(Monet et al. \cite{monet96}) and the ROE/NRL catalog (Yentis et al. 
\cite{yentis92}),  a precision around 1\,arcsec being achieved. The 
spatial distribution of the candidates within the area surveyed is 
shown in Fig.~\ref{fig1} and finder charts in the $I$-band 
(3\arcmin$\times$3\arcmin \ in extent) are provided in Fig.~\ref{charts}.  

\subsection{Spectroscopy}
We have obtained low-resolution  optical spectroscopy of nine of our 
photometric candidates (SOri\,12, 17, 25, 27, 29, 39, 40, 44, and~45), 
using the 4.2-m William Herschel Telescope (WHT) at the ORM. The spectra 
were collected on 1997 December 28--30;  the instrumentation used was 
the ISIS double-arm spectrograph (the red arm only), the R158R grating 
and the TEK 1024$\times$1024\,pixel$^2$ CCD detector which provides a 
total spectral coverage of 635--920\,nm and a nominal dispersion of 
2.9\,\AA\ per pixel.  The spectral resolution (FWHM) of the instrumental 
setup was 20\,\AA .  Exposure times ranged from 900 to 2500\,s depending 
on the magnitudes of the objects and on the weather conditions. Spectra 
were reduced by a standard procedure using IRAF, which included debiasing, 
flat-fielding, optimal extraction,  and wavelength calibration using the 
sky lines appearing in each individual spectrum (Osterbrock et al. 
\cite{osterbrock96}).  Finally, the spectra were corrected for the 
instrumental response making use of the standard star G\,191-B2B, which 
has absolute flux data available  in the IRAF environment.  

All the spectra clearly correspond to late M-type objects, showing typical 
VO and TiO molecular bands. We have classified them by comparison to 
Pleiades members of known spectral types (Mart\'\i n et al. \cite{martin96}; 
Zapatero Osorio et al. \cite{osorio97b}) resulting that our objects range 
from class M6 to M8.5. We have also obtained the pseudo-continuum PC1--4 
indices (Mart\'{\i}n et al. \cite{martin96}) and found them to yield 
slightly earlier spectral types by about half a subclass with respect to 
the main sequence field dwarfs. This is likely due to an effect of gravity; 
nevertheless this difference is within the estimated error bar of our 
measurements. The spectra of eight of the candidates (all except for 
SOri\,44) are shown in Fig.~\ref{fig4}, where the clearest features are 
indicated.

\section {Discussion}
\subsection{Contamination by other sources}
Possible contaminating objects in our survey are red galaxies, M giants, 
and foreground field M dwarfs. Due to the spatial resolution and 
completeness magnitudes of our observations,  the contamination due 
to red galaxies is not a major problem, since  these are mostly resolved 
and routines in IRAF can distinguish them from stellar point-like objects. 
Another source of contamination is the M giants, but given the galactic 
latitude of the cluster ($b$\,=\,--17.34\,deg) their number is negligible 
($\le 5 \%$) in comparison with main-sequence dwarfs  (Kirkpatrick et al. 
\cite{kirk94}).

The most relevant source of contamination is field M dwarf stars in the 
line of sight  towards the cluster.  To estimate their number we have 
considered the results  from searches covering a large area of the sky. 
Kirkpatrick et al. (\cite{kirk94}) performed a 27.3\,deg$^{2}$ survey, 
reaching a completeness magnitude of $R$\,=\,19\,mag and finding space 
densities of 0.0022\,pc$^{-3}$,  0.0042\,pc$^{-3}$,  and 0.0024\,pc$^{-3}$ 
for M5--M6, M6--M7, and M7--M9 dwarfs, respectively.  With these densities 
and with the typical absolute magnitudes of late M dwarfs (Kirkpatrick 
\& McCarthy \cite{kirkmc94}) we have calculated the number of these cool 
stars that might be populating the CM region which  we ascribe to the 
cluster members. The result is that less than one M5--M6, about one M6--M7, 
and one M7--M9 field  dwarfs should be contaminating  our survey. Our 
selection criterion based on the three filters $R$, $I$, and $Z$ appears to 
be good enough to differentiate clearly between true members and 
contaminating field stars, but further studies in the near infrared, which 
is less affected by reddening, or low-resolution spectroscopy will tell 
us which objects are definitely bona fide cluster members.

\subsection{The $\sigma$\,Orionis cluster: size, age, and distance}

The existence of a cluster around the multiple star $\sigma$\,Orionis 
was noted for the first time  in the Lund Observatory Catalogue of Open 
Clusters (Lynga \cite{lynga81}), where it was designated by the name 
of the star. In later studies (Lynga \cite{lynga83}) about fourteen 
stars were given as members and the diameter of the cluster in the sky 
was estimated at 25\,arcmin. The work by Wolk (\cite{wolk96}) and Walter 
et al. (\cite{walter97})  covered an area of 900\,arcmin$^2$ containing 
a rather  dense population of X-ray sources,  and found a homogeneous 
distribution of cluster candidates. In the future, a larger area will 
need to be surveyed to determine with precision  the total region 
occupied by the cluster, since our current knowledge may be limited 
to the core. 

Age is one of the most important parameters of a cluster, particularly 
in locating the substellar mass limit. Until recently, the only way 
to determine the age of $\sigma$\,Orionis was via studies of the massive 
stars of the OB1b subgroup, resulting in age estimates in the range of 
1.7--7\,Myr,  (Blaauw \cite{blaauw91}; WH; BGZ). The discovery by Wolk 
(\cite{wolk96}) of a large low-mass population allowed him to compare 
H--R diagrams  with theoretical isochrones and  obtain an age  for the 
cluster of about 2\,Myr (Wolk \& Walter \cite{wolk99}), in good agreement  
with  estimates for the subgroup, and especially with the age given by 
the latest work of BGZ (1.7\,Myr).  This provides additional support for  
the assumption that the cluster belongs to the young Orion star forming 
region, and that its central star, $\sigma$\,Orionis, is indeed a member 
in the cluster of the sa{\rm }me name. Figure~\ref{fig5} shows our candidates 
together with theoretical isochrones from several authors ({\sl (a)} 
Burrows et al. \cite{burrows97}; {\sl (b)} D'Antona \& Mazzitelli 
\cite{dantona97}; {\sl (c)} Baraffe et al. \cite{baraffe98}). The later 
models provide magnitudes and colors and in order to transform the 
effective temperatures and luminosities of the models by Burrows et al. 
(\cite{burrows97}) and D'Antona \& Mazzitelli (\cite{dantona97}) into the 
observational CM diagrams of Fig.~\ref{fig5}, we have used the 
temperature--color scale and bolometric corrections from Bessell, Castelli 
\& Plez (\cite{bessell98}). The cluster candidates seem to imply average 
ages of $\le$1\,Myr to 3\,Myr accor{\bf }ding to Burrows et al. 
(\cite{burrows97}), and $\le$1\,Myr to 5\,Myr according to D'Antona 
\& Mazzitelli \cite{dantona97}). The evolutionary tracks by Baraffe et al. 
(\cite{baraffe98}) used in Fig.~\ref{fig5}~{\sl (c)} are those named 
``dusty'' models by these authors and they include dust condensation and 
opacity in the atmospheres of cool objects. While these models predict 
effective temperatures and luminosities very similar to the other two 
sets of isochrones, the predicted magnitudes and colors do not fit the 
observations. Alternative methods for deriving ages could give a more 
precise age for $\sigma$\,Orionis. In particular, the Li-luminosity 
(LL-clock) dating (see e.g. Mart\'{\i}n \& Montes \cite{martin97};  Basri 
\cite{basri98}), which provides a very good age determination in the 
case of the Pleiades and $\alpha$-Persei clusters (Mart\'{\i}n et al. 
\cite{martin98b}; Stauffer, Schultz \& Kirkpatrick \cite{stauffer98}; 
Basri \& Mart\'\i n \cite{basri99}) can be useful in this case as it 
will be discussed in section~\ref{secLi}.

The distance is another important parameter that needs to be known for 
a cluster.  Measurements available in the literature give a distance 
modulus of 7.8--8 (WH; BGZ) for the subgroup. More recently,  
{\it Hipparcos} has provided a distance to the central star of the 
cluster $\sigma$\,Orionis of  352\,pc ($m - M$=7.73), slightly smaller 
than previous results, but in good agreement with them. We have adopted 
this value as the cluster distance.

\subsection{Spectroscopy of brown dwarf candidates}
From the nine objects studied spectroscopically, eight appear to be very 
probable young cluster members, which implies a high efficiency 
(89$\%$) for our photometric search strategy. All eight confirmed 
candidates (listed in Table~\ref{tab4}) show spectral types later than 
M5 (three M6, two M6.5, two M7, and one M8.5). The rejected candidate 
(SOri\,44) is hotter (M6.5) than expected for its given $I$ magnitude 
and does not show spectral features indicative of the youth of the 
$\sigma$\,Orionis cluster, such as H$\alpha$ in emission. In 
Table~\ref{tab4} we give the strengths of the H$\alpha$ and 
Na\,{\sc i} lines. In the Pleiades and 
$\alpha$~Persei clusters the M6.5 spectral type marks the  substellar 
mass limit (Mart\'{\i}n et al. \cite{martin98b}; Basri \& Mart\'\i n 
\cite{basri99}), with an  uncertainty of about half a subclass. As very 
low-mass stars and massive brown dwarfs evolve at nearly constant 
temperature ($\pm$200\,K) from several million years to nearly 
100\,Myr (Baraffe et al. \cite{baraffe98}; Burrows et al. \cite{burrows97}; 
D'Antona \& Mazzitelli \cite{dantona94}), we expect that the substellar mass 
limit is located at a similar spectral type for ages of a few Myr 
(Luhman et al. \cite{luhman98a}). This should be taken with caution 
if the relationship between effective temperature and spectral type for 
low-gravity objects is different from that of dwarfs or 100\,Myr old objects.
All our spectroscopically confirmed candidates with spectral types later than 
M7 are therefore very likely brown dwarfs. In Fig.~\ref{fig6} we 
can see that the eight objects show indications of strong activity, with 
higher H$\alpha$ emission than Pleiades objects of the  same spectral 
type. This argues in favor of a younger age, since it is expected that 
activity decreases with the age. We also see variations in the emission 
of objects of similar type, a kind of behavior already seen in late M 
objects of young clusters like the Pleiades, IC\,348 or Taurus 
(Zapatero Osorio et al. \cite{osorio97b}; Luhman et al. 
\cite{luhman98b}; Brice\~{n}o et al. \cite{briceno98}). Spectral features 
associated with Na\,{\sc i} and K\,{\sc i} can be seen in some spectra, 
while in others they are too weak and we can only set upper limits. The 
smaller equivalent width (EW) of Na\,{\sc i} in our $\sigma$\,Orionis 
candidates (except for SOri\,44 which shows an absorption typical of 
ages older than 100\,Myr) with respect to Pleiads and older field objects 
of the same spectral type may be a result of the lower surface gravity of 
these very young objects (Mart\'{\i}n et al. \cite{martin96}; Luhman et al. 
\cite{luhman97}; Brice\~{n}o et al. \cite{briceno98}).

\subsection{The coolest brown dwarf in the $\sigma$\,Orionis cluster}
Our candidate of latest spectral type (M8.5), SOri\,45, deserves special 
attention  since it is among the coolest and therefore among the least 
massive objects in our sample. This very young brown dwarf candidate has an 
effective temperature in the range between 2100 and 2500\,K, as derived 
from different temperature scales for dwarfs available in the 
literature (Tinney et al. \cite{tinney93}; Kirkpatrick \cite{kirk95_1}; 
Jones et al. \cite{jones96}; Leggett et al. \cite{leggett98};  
Bessell et al. \cite{bessell98}). However, we note that the temperature 
scale for giants is several hundred degrees warmer and an upward correction 
in the estimated temperature of about 100--200\,K (Luhman et al. 
\cite{luhman97}) may be required. The luminosity of the object can be 
obtained from the $I$ magnitude using the bolometric corrections  of various 
authors (Monet et al. \cite{monet92}; Tinney et al. \cite{tinney93}; 
Kirkpatrick, Henry \& Simons \cite{kirk95};  Bessell \& Stringfelow 
\cite{bessell93}; Bessell et al. \cite{bessell98}). We derived an average 
correction factor $BC_{\rm I} =- 1.1 \pm 0.1$. From a comparison of the 
colors of our objects with those of objects of the same spectral type in 
the Pleiades and field stars, we do not find any significant reddening 
($A_{V} \le 0.5$\,mag). So, assuming that the extinction is negligible, 
and taking as distance modulus of the cluster $m - M$~=~7.73, we obtain a 
luminosity for our object of $\log L/L_{\odot} = - 2.40 \pm 0.15$. If we 
adopt 10\,Myr as an upper limit for its age, according to theoretical 
calculations of luminosities for young ages (Baraffe et al. 
\cite{baraffe98}; D'Antona \& Mazzitelli  \cite{dantona97}; Burrows et al. 
\cite{burrows97}) we infer a mass of 0.020--0.040\,$M_{\odot}$ for 
SOri\,45, and for the most likely age of the cluster, 2--5\,Myr, a mass 
of only 0.020--0.025\,$M_{\odot}$. Therefore it is one of the least 
massive objects found to date outside the Solar System. Follow-up 
IR and/or spectroscopic observations may reveal even less massive objects 
among the faintest $\sigma$\,Orionis cluster photometric candidates of 
Table~\ref{tab3}. The uncertainties in the 
conversion from magnitudes to luminosities and in the theoretical 
modeling  of such low-mass objects at very early ages are considerable,  
hence this mass estimate is to be treated with caution. Nevertheless, 
our results leave little doubt concerning the extension of the star 
formation process very deep in the brown dwarf domain. In Table~\ref{tab4} 
luminosities and masses are derived using the same procedure given above 
for the remaining of the eight candidates for which spectra were obtained.  

In Fig.~\ref{fig7} we compare our least massive brown dwarf candidate of 
Table~\ref{tab2}, SOri\,45, with other objects of  spectral type M8.5 
from Ophiuchus, the Pleiades and the field. The stronger H$\alpha$ 
emission and the much lower strength  of K\,{\sc i} and Na\,{\sc i} 
lines are noteworthy (Figs.~\ref{fig7}(b, c, d)). As mentioned above, 
this is to be expected for  a very young object. Another interesting 
feature is that the VO and TiO molecular bands are clearly less intense 
in the field  star than in SOri\,45 (see the spectral regions 660--760\,nm 
and 840--880\,nm, Fig.~\ref{fig7}(c)) .  This is possibly a consequence of 
the higher gravity of the older systems, which favors the condensation 
of dust grains in cold atmospheres (Tsuji, Ohnaka, \& Aoki \cite{tsuji96}). 
In Fig.~\ref{fig7}(a) we show a comparison with the spectrum of the M8.5 
object found by Luhman et al. (\cite{luhman97}) in Ophiuchus.  The 
similarity of the two spectra is remarkable.  These two objects, although 
discovered in star forming regions far away from each other, appear to 
be extremely similar, suggesting that low-mass brown dwarfs could be quite 
common. Old counterparts of these substellar objects may be populating the 
galactic disk. At the age of a few Gyr their atmospheric temperatures will 
be similar to or likely lower than that of Gl\,229B ($\sim$ 1000\,K), 
so it is expected 
that they present spectroscopic characteristics intermediate between 
those of Gl\,229B and Jupiter. Important questions that remain unanswered 
are how many objects of this kind there are, and whether even less 
massive ones can form. These questions are obviously related to our 
knowledge of the  mass function at such low masses, but beyond the scope 
of this article. Although we believe that the area covered and the number 
of objects are significant, at this juncture this issue presents several 
difficulties mainly due to the uncertainty in the cluster age; small 
changes in the adopted age imply a significantly different mass-luminosity 
relationship and consequently, large variations in the slope of the mass 
function. To this difficulty we must add the uncertainty in the  
membership of our candidates, especially the faintest ones, which  
might be more contaminated by reddened field stars. Nevertheless, if we 
adopt the age (about 3\,Myr) of the model which provides the best fit to 
the optical photometric sequence of $\sigma$\,Orionis we derive that the 
number of brown dwarfs per unit of mass grows through the substellar 
domain till around 0.040\,$M_{\odot}$, which is in agreement with what 
is seen in the Pleiades (Mart\'{\i}n et al. \cite{martin98c}; Bouvier 
et al. \cite{bouvier98}).

\subsection{\label{secLi} Lithium and deuterium depletion: prospects of 
identification of substellar objects and age determination}
Young stellar clusters such as $\sigma$\,Orionis are excellent 
laboratories for studying the  depletion  of light elements such as 
deuterium, lithium, berillium, and boron, since these elements are burned 
in the early stages of  pre-main sequence stellar evolution and their 
atmospheric abundances drastically change in this phase. As mentioned before, 
a detailed knowledge of lithium burning at the bottom  of the main sequence 
has  provided an alternative method of age determination for the young 
clusters $\alpha$~Persei and the Pleiades. This method appears to be more 
reliable than traditional ones, based on evolutionary models of the more 
massive stars in the clusters  which are limited by our poor knowledge 
of the interiors of these stars.  Lithium dating relies on the fact that 
in fully convective low-mass stars lithium burning takes place over a 
very short time interval (a few Myr)  once the temperature in the core 
is high enough to produce  the destructive reactions (Li,p)$\alpha$. The 
lower the mass of the star  the greater is the age at which it starts 
lithium burning; in any case this age is always smaller than a few tens 
of Myr. Brown dwarfs less massive than 0.060--0.065\,M$_\odot$ do not 
ever burn lithium because their  cores do not reach the minimum burning 
temperature. The presence of lithium in the atmosphere of low 
luminosity, fully convective objects is a clear indication that the 
maximum internal temperature is below the lithium burning value, and  that 
the object is sub-stellar. However this simple criterion can  be applied 
only if the object is older than 150\,Myr. Those objects with masses above 
0.065\,M$_\odot$ do destroy lithium while they are young ($\le$150\,Myr) 
and this fact can be used for dating clusters. 

A detailed inspection of evolutionary models confirms  that the transition 
between lithium depletion/preservation in the atmospheres of low-mass 
objects takes place at higher masses and luminosities as we consider 
younger ages. Given the youth of $\sigma$\,Orionis, this transition should 
occur in early/mid M-type stars. If the age were indeed 10\,Myr we would 
expect  stars in the mass range 0.9--0.4\,M$_\odot$ 
($-1.0 \le \log L/L_\odot \le -0.5$) to have  burned most of their original 
lithium content, whereas if the cluster is as young as 2--5\,Myr, neither 
stars nor massive brown dwarfs would have had sufficient time to reach 
the   interior  temperatures needed to start lithium burning (D'Antona 
\& Mazzitelli \cite{dantona97}; Soderblom et al. \cite{soderblom98}).  
It is consequently very important to search for lithium in the early/mid 
M-type stars of the cluster. 

{\sl The Deuterium Test}\\
Deuterium has a similar behavior to that of Li as it is a light element 
that is destroyed in stellar interiors when the temperature reaches 
0.8$\times$10$^6$\,K (Ventura \& Zeppieri \cite{ventura98}), i.e., much 
below the minimum temperature for lithium 
burning. Therefore, deuterium burning takes place at much earlier ages 
and extends to less massive substellar objects (0.015--0.018\,$M_{\odot}$, 
Burrows et al. \cite{burrows93}; D'Antona \& Mazzitelli \cite{dantona97}; 
Burrows et al. \cite{burrows97}). Brown dwarfs with masses larger 
than 0.02\,M$_\odot$ efficiently destroy deuterium in the age range 
1--10\,Myr and this destruction takes place on a very short time scale. 
Figure~\ref{fig8} represents the evolution of the deuterium abundance as 
a function of age for several masses. Only  objects below 0.02\,M$_\odot$ 
can preserve their original abundance on time scales of 10\,Myr, since it 
will take longer for them  to burn any deuterium. During the 
deuterium-burning phase the luminosity and effective temperatures stay 
almost constant (Burrows et al. \cite{burrows93}; D'Antona \& Mazzitelli 
\cite{dantona94}). All the physical arguments discussed above for lithium 
can be applied to deuterium. There is a transition between objects which 
have burned deuterium and those which have preserved it which  in principle 
can provide an age determination method. In Fig.~\ref{fig8}, we can note, 
as an example, how an object of 0.07--0.08\,M$_\odot$ with an age of 
3\,Myr has destroyed a significant amount of its initial deuterium 
abundance, whereas a 0.03\,$M_\odot$ object does not burn deuterium (by 
a factor larger than 10) until an age of 7\,Myr. If $\sigma$\,Orionis were 
as young as 2--3\,Myr the transition between deuterium burning and 
preservation would take place at $\log L/L_\odot=-1.6$ which approximately 
coincides with the substellar mass limit (0.070--0.075\,$M_{\odot}$), 
while at an age of 10\,Myr the deuterium preservation would take 
place at $\log L/L_\odot \sim -3$ (D'Antona \& Mazzitelli 
\cite{dantona97}). We have considered theoretical models with the 
interstellar abundance of deuterium (see e.g. Linsky \cite{linsky98}) for 
this discussion. If the initial abundance of this isotope is changed by 
50\%, the time scale for the deuterium depletion is subsequently affected 
by a factor $\sim$\,1.5. For a given age the lower the abundance of 
deuterium is, the lower the luminosity and the mass at the deuterium 
depletion/preservation boundary.

It is interesting to note that at the approximate age of 4\,Myr, all 
low-mass stars will have destroyed their deuterium by several orders of 
magnitude; therefore, the simple detection of this isotope  in an older 
object would manifest its substellar nature. The detection of deuterium, 
then, complements that of lithium as a  test of substellarity for young 
objects and provides a way of confirming the substellar nature of objects 
which, due to their young ages, cannot be confirmed as such  by the 
lithium test. We shall also note an alternative way to make use of the 
potential of deuterium observations: any object older than 1\,Myr, with 
a luminosity below $\log L/L_\odot=-1.5$ and where deuterium is preserved, 
must be substellar. 

The detection of deuterium is difficult and presents an important 
challenge from the observational point of view. Deuterium has been 
detected in the planets and comets of the Solar System among other 
astrophysical sites. The observations were made using dipole lines of 
monodeuterated hydrogen  in the visible  (Macy \& Smith \cite{macy78}) 
and rotational bands of monodeuterated methane at 1.6\,$\mu$m (De Bergh 
et al. \cite{deberg86}, \cite{deberg88}, \cite{deberg90}). Implications 
on primordial deuterium could be deduced if detection were achieved in 
young brown dwarfs, or in those less massive than 0.015\,$M_{\odot}$. 
The very low-mass $\sigma$\,Orionis cluster members are in the phase of 
burning this light element. Figure~\ref{fig9} shows our candidates in 
an H--R diagram, which includes the frontier between the depletion (by 
a factor 10) and preservation of deuterium for different ages. To convert 
$I$ magnitudes and colors to luminosities and temperatures, we have taken 
account of the same references given in \S 3.4. As we can see, if the 
cluster were indeed as young as it has been claimed (around 3\,Myr),  
all objects with luminosities below log\,$L/L_{\odot}$\,=\,$-1.7$ and 
temperatures cooler than $\sim$2700\,K (about M6--M7 spectral types) could 
preserve their deuterium, while if the cluster were as old as 10\,Myr, 
this could happen only to those with the latest types. Objects like 
SOri\,45 are then the most promising candidates for detecting deuterium. 
Observations of deuterium in a substantial number of $\sigma$\,Orionis 
members could also be used to constrain any age spread within 
the cluster.

\section{Conclusions}
We have performed a deep $R$, $I$, and $Z$ survey in the very young 
cluster $\sigma$\,Orionis covering an area of 870\,arcmin$^2$ and have 
found objects with masses as small as 0.020--0.025 $M_{\odot}$, well 
below the substellar mass limit. Our selected 49~candidates 
define a photometric sequence ranging in age from $\le$1\,Myr to 5\,Myr, 
which is in agreement with previous results for the OB1b subgroup, 
where the multiple star $\sigma$\,Orionis is associated. 

We have confirmed spectroscopically the cool nature of eight of these 
objects (spectral types M6--M8.5), which show spectral features indicative 
of a stronger activity and lower gravity than previously known members of 
similar types in older clusters. Our latest candidate, SOri\,45 
(M8.5), is one of the least-massive objects known to date, with a best 
estimate of its mass at 0.020--0.025\,$M_{\odot}$ for the age of 
2--5\,Myr. An upper limit of 0.040\,$M_{\odot}$ is estimated  
if the age of the cluster were as old as 10\,Myr. The detection of the 
old counterparts of these brown dwarfs in the solar neighborhood will 
represent a challenge for future searches as they will cool down to very 
low atmospheric temperatures ($T_{\rm eff} \la 1000$\,K). At the age 
of the Sun a brown dwarf of 0.020\,$M_{\odot}$ may have absolute magnitudes 
of around  $M_{J}$ $\sim$ 20--21, and $M_{K}$ $\sim$ 22--23\,mag 
(Burrows et al. \cite{burrows97}). Given the limiting magnitudes of 
the current large scale infrared surveys like DENIS (Delfosse et al. 
\cite{delfosse99}) and 2MASS (Beichman et al. \cite{beichman98}), the 
detection of such low-mass brown dwarfs would be possible if they were 
located at distances closer than about 3\,pc.

The study of light elements like lithium and deuterium in brown dwarfs 
of young clusters may provide a precise tool of determining ages. In 
particular, the least-massive brown dwarfs that we have found in 
$\sigma$\,Orionis should have preserved their deuterium and it is worthwhile 
investigating possible ways of achieving its detection.

\acknowledgments
{\it Acknowledgments}: We thank A. Oscoz for acquiring data at the 
IAC-80 Telescope necessary for the calibration photometry of the WFC 
observations. We thank  K. Luhman for kindly have provided the M8.5 
spectrum in $\rho$ Ophiuchi, and I. Baraffe and the Lyon group  and   
F. D'Antona for sending us electronic versions of their recent models. This 
work is based on observations obtained at the  INT and WHT  operated by 
the Isaac  Newton Group of Telescopes funded by PPARC at the Spanish 
Observatorio del Roque de los Muchachos of the Instituto de 
Astrof\'{\i}sica de Canarias and the IAC80 Telescope at the Observatorio 
del Teide (Tenerife, Spain).  Partial financial support was provided 
by the Spanish DGES project no. PB95-1132-C02-01.

\clearpage

\figcaption[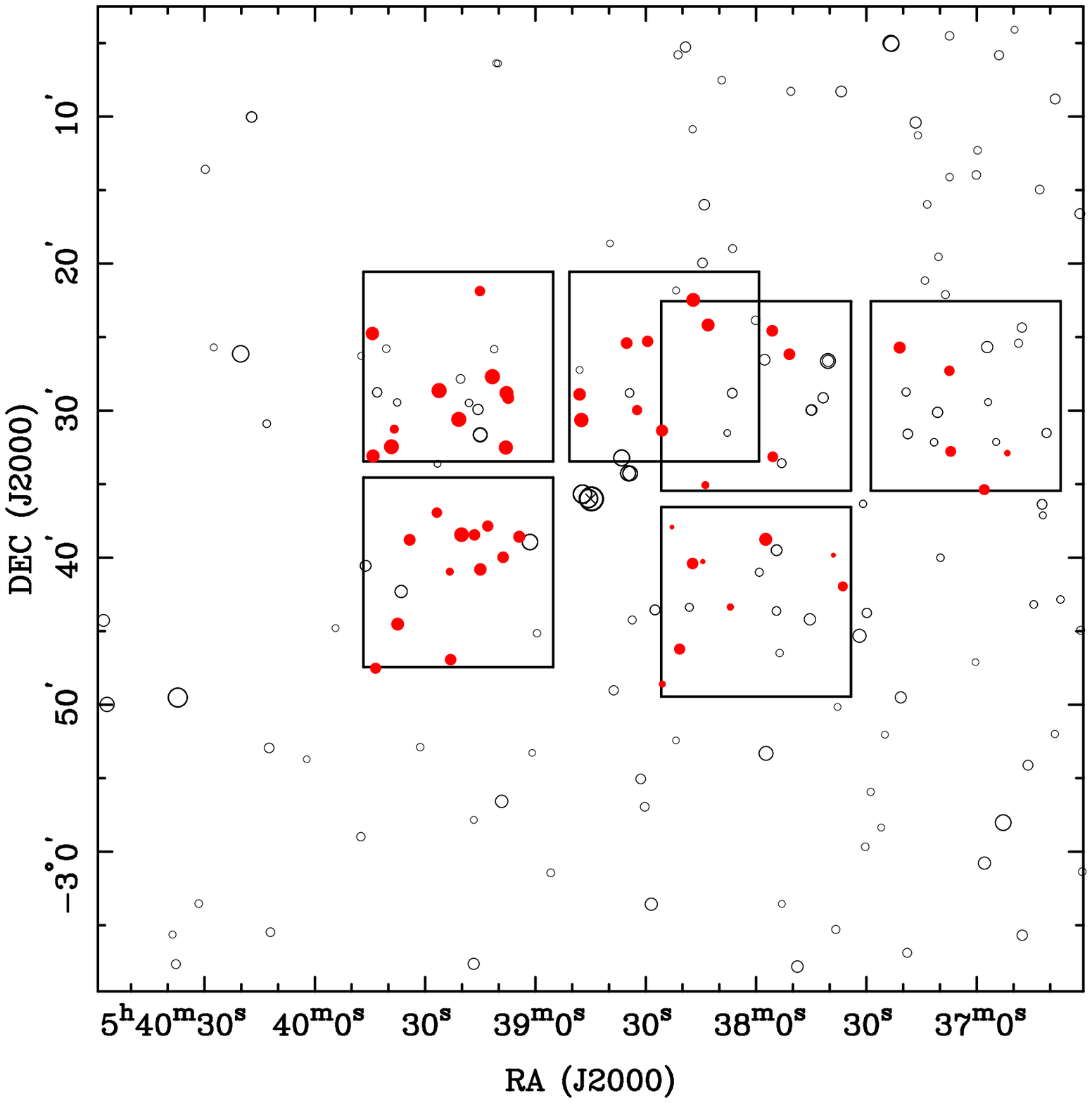]{\label{fig1} Location of our fields (open squares) 
covering an area of 870\,arcmin$^2$  around the multiple star 
$\sigma$\,Orionis (indicated with a cross). Filled symbols represent our 
candidates (Tables~2 and 3) and open circles denote those stars in the 
field brighter than 13\,mag. The relative brightness is denoted by 
circle diameters. North is up and East is left.}

\figcaption[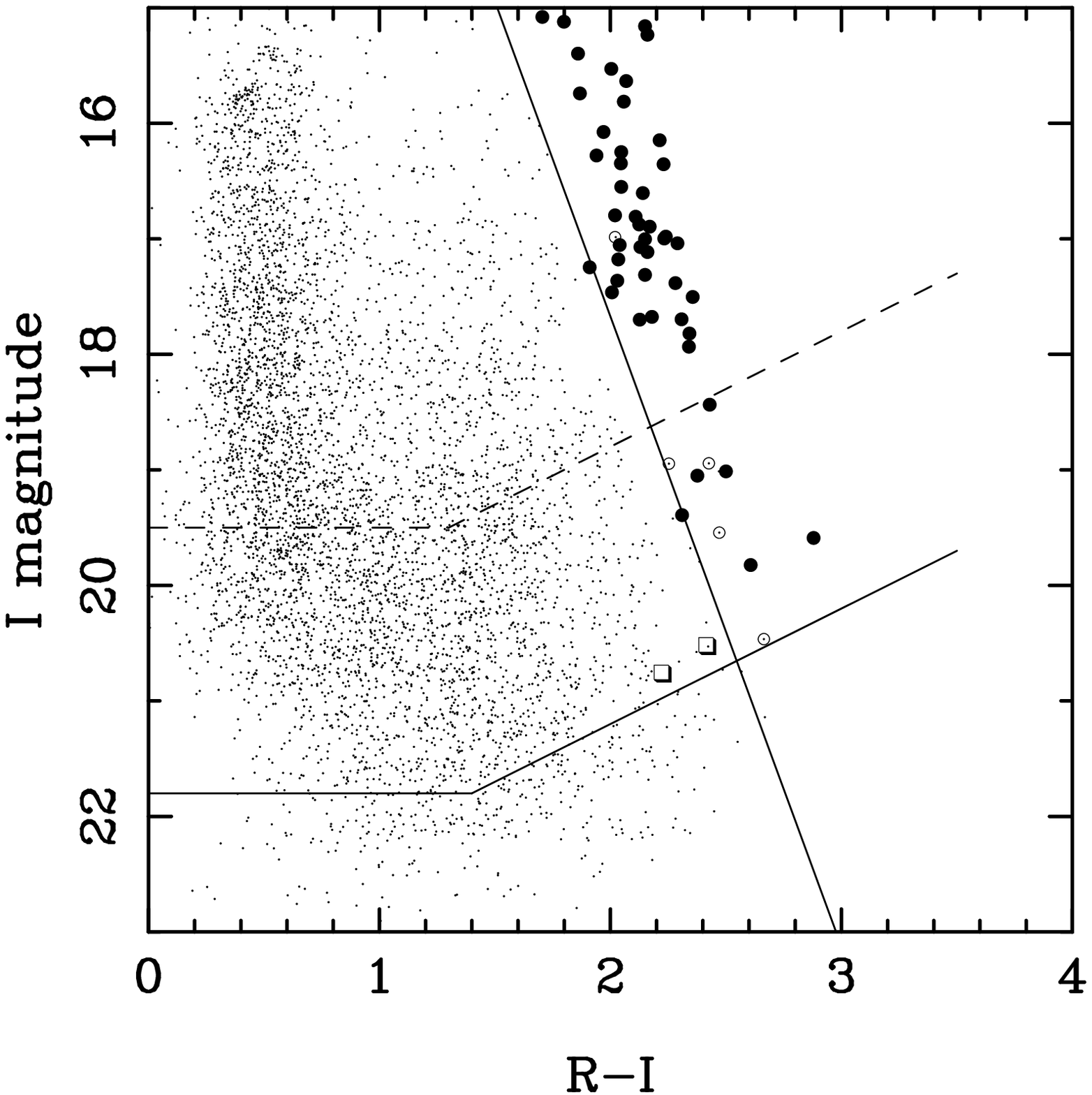]{\label{fig2} Color-magni\-tude diagram for very 
low-mass stars and brown dwarfs in the $\sigma$\,Orionis cluster resulting 
from our survey. Filled circles denote our 46 candidates listed in Table~2. 
Those objects that were not found red in all colors are indicated with 
open symbols; the ones with $R-I$ measured in Table~3 are shown with open 
squares. The straight, vertical line separates cluster member candidates 
from field objects (see text). Completeness and limiting magnitudes are 
indicated with a dashed line and a full line, respectively.}

\figcaption[bejar3a.eps,bejar3b.eps]{\label{charts} Finder charts in the 
$I$-band (3\arcmin$\times$3\arcmin \ in exent). North is up and East is 
left.}

\figcaption[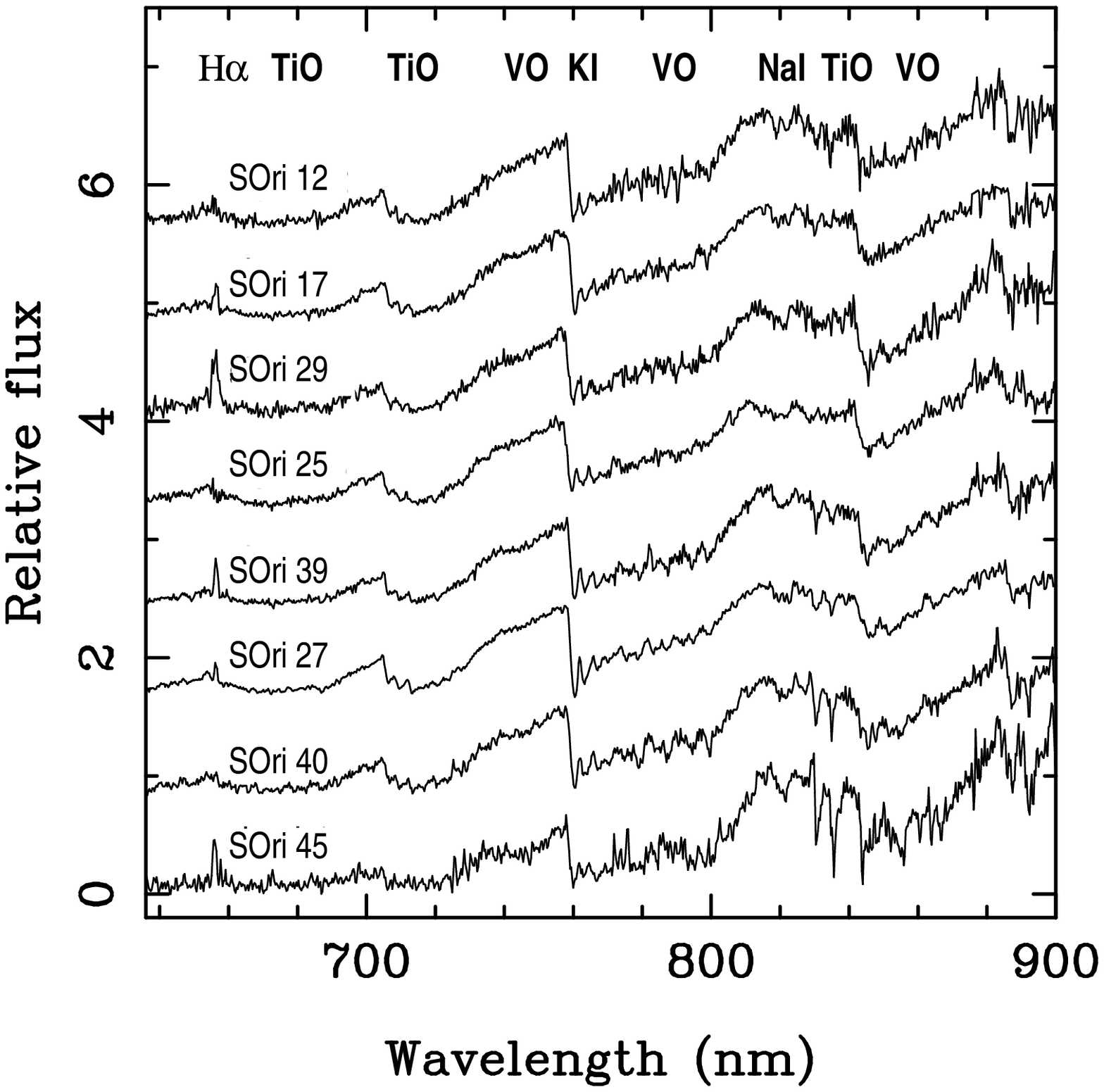]{\label{fig4} Low-resolution (20\,\AA) optical 
spectra of eight of our candidates normalized at 813\,nm. A constant 
step of 0.8\,units has been added to each spectrum for clarity.}

\figcaption[bejar5.ps]{\label{fig5} Color-magnitude diagram for our 
$\sigma$\,Orionis candidates compared with the recent evolutionary tracks of 
{\sl (a)} Burrows et al. (1997), {\sl (b)} D'Antona \& Mazzitelli (1997), 
and {\sl (c)} Baraffe (1998). Ages and masses in solar units are indicated. 
Filled circles denote the candidates listed in Tables~2 and~3, and those 
objects with spectra presented in this paper are indicated with open 
triangles. Bright stars ($I \le 15$\,mag, open circles) are taken from 
Wolk (1996), those ones which are confirmed spectroscopically. }

\figcaption[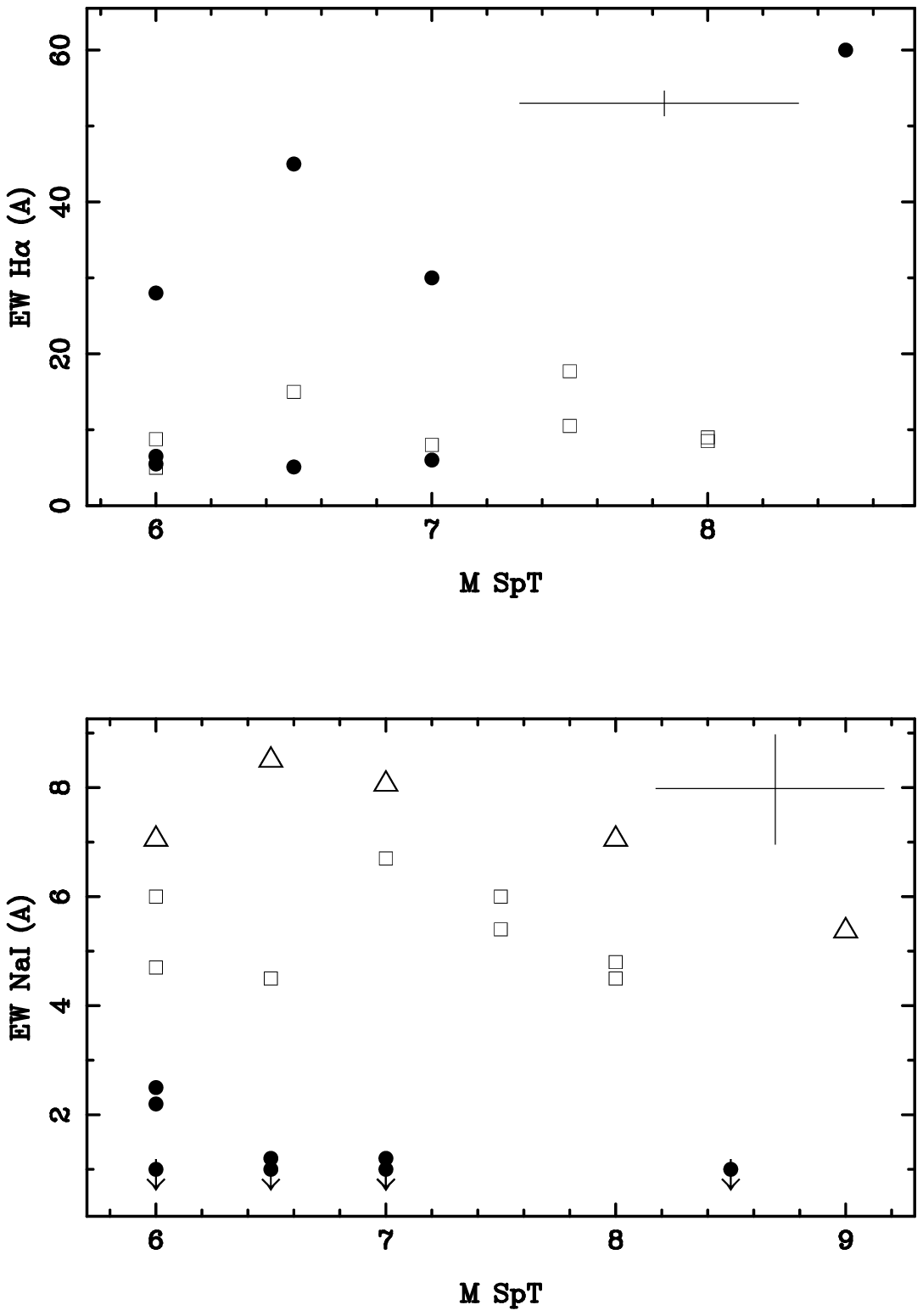]{\label{fig6} The Na\,{\sc i} and H$\alpha$ 
equivalent widths measured in our $\sigma$\,Orionis candidates (filled 
circles), in some Pleiades members (open squares) and in some field stars 
(open triangles) plotted against spectral type. Error bars of our 
measurements are indicated in the upper right corner.}

\figcaption[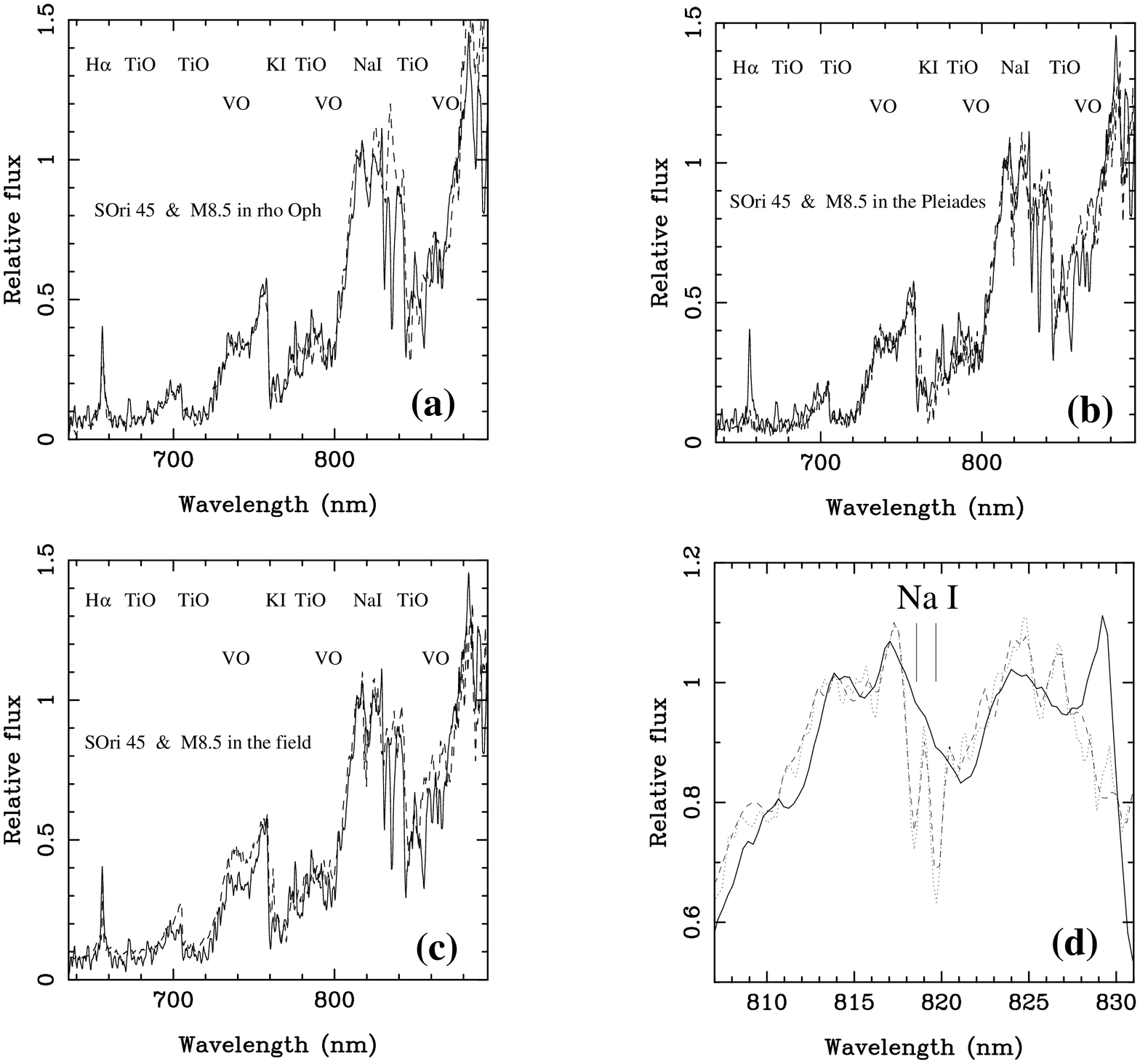]{\label{fig7} Our candidate SOri\,45 (M8.5 spectral 
type, full line) compared to its spectral type counterpart (dashed line) 
in {\sl (a)} Ophiuchus (Luhman et al. 1997), {\sl (b)} the Pleiades, and 
{\sl (c)} the field. The Pleaides spectrum has been obtained averaging 
those of the brown dwarfs Teide\,1 (M8, Mart\'\i n et al. 1996) and 
Roque\,4 (M9, Zapatero Osorio et al. 1997b). In panel {\sl (d)} an 
enlargement of the Na\,{\sc i} doublet is presented where SOri\,45 is 
denoted with a full line, and the Pleiades and field M8.5-type spectra 
are plotted with dashed and dotted lines, respectively.}

\figcaption[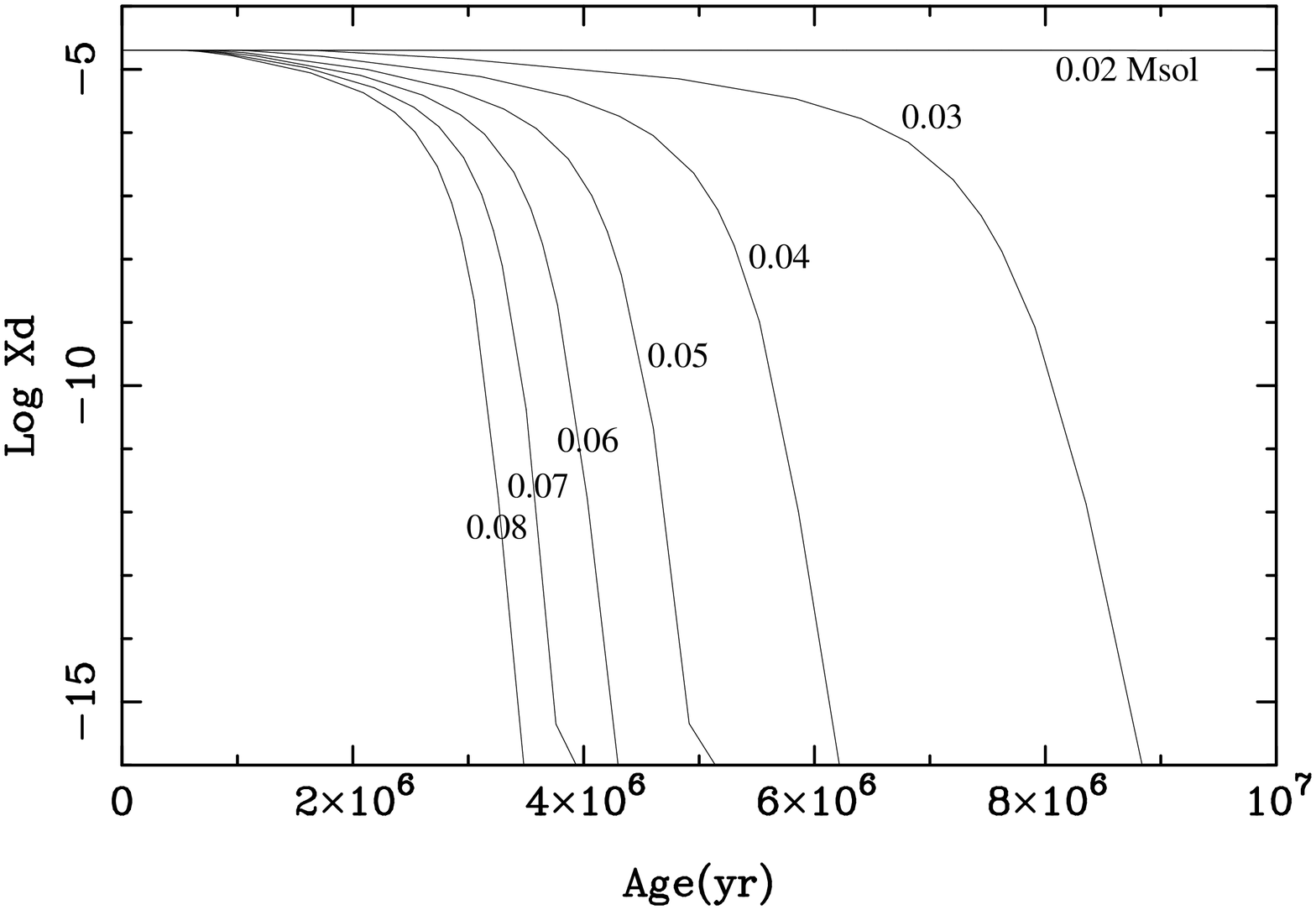]{\label{fig8} The deuterium burning as a function 
of age (models of D'Antona \& Mazzitelli 1998, those with the 
interestellar abundance of deuterium). Masses are indicated in solar units.}

\figcaption[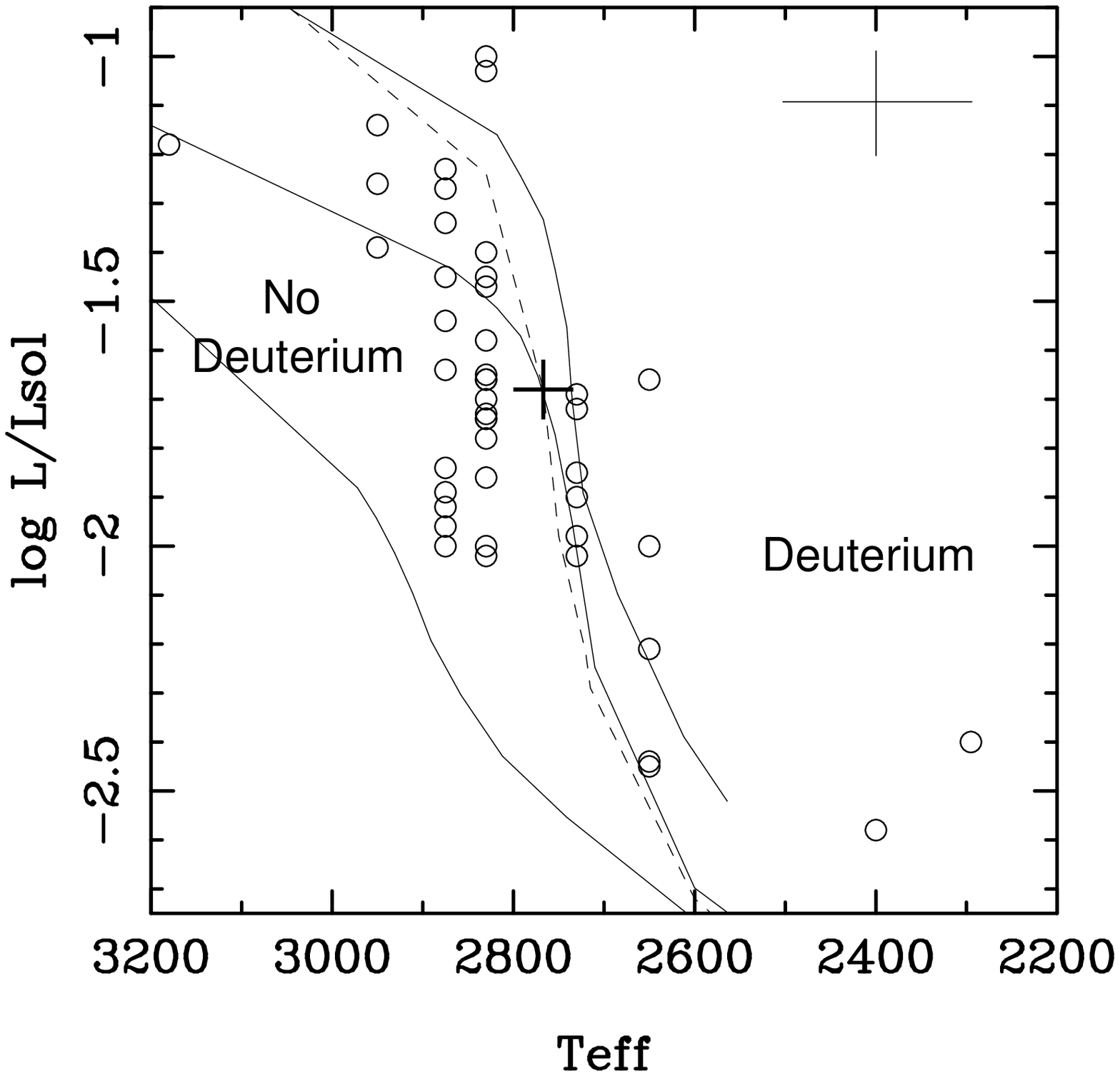]{\label{fig9} The HR diagram for our $\sigma$\,Orionis 
candidates (open circles). D'Antona \& Mazzitelli's (1998) theoretical 
isochrones of 1, 3 and 10\,Myr are superimposed. The borderline for the 
deuterium depleted by one order of magnitude (90\% \ preservation, and 
interstellar initial deuterium abundance) is indicated with the dashed 
line. The cross symbol stands for the location of this borderline at 
the age of 3\,Myr ($\sim$0.060\,$M_{\odot}$).}

\clearpage

\begin{table}\begin{center}
\caption{\label{tab1} Completeness and limiting magnitudes}
\begin{tabular}{ccccccccc}
 & & & & & & & & \\
\hline
\hline
\multicolumn{3}{c}{} &
\multicolumn{3}{c}{Completeness} &
\multicolumn{3}{c}{Limiting} \\
\multicolumn{3}{c}{} &
\multicolumn{3}{c}{-----------------------} &
\multicolumn{3}{c}{-----------------------} \\
\multicolumn{1}{c}{Fields} &
\multicolumn{1}{c}{R.A.} &
\multicolumn{1}{c}{Decl.} &
\multicolumn{1}{c}{$R$} &
\multicolumn{1}{c}{$I$} &
\multicolumn{1}{c}{$Z$} &
\multicolumn{1}{c}{$R$} &
\multicolumn{1}{c}{$I$} &
\multicolumn{1}{c}{$Z$} \\
\hline
1 & 5 39 21 & --2 27 00 & 20.5 & 19.5 & 19.0 & 21.8 & 20.8 & 20.2\nl
2 & 5 38 25 & --2 27 00 & 20.5 & 19.5 & 19.0 & 21.8 & 20.8 & 20.2\nl
3 & 5 39 21 & --2 41 00 & 20.2 & 19.2 & 19.2 & 21.2 & 20.2 & 20.2\nl
4 & 5 37 03 & --2 29 00 & 22.5 & 20.7 & 20.2 & 23.8 & 22.2 & 21.5\nl
5 & 5 38 00 & --2 43 00 & 21.0 & 20.2 & 19.7 & 22.2 & 21.0 & 20.5\nl 
6 & 5 38 00 & --2 29 00 & 22.5 & 20.7 & 20.2 & 23.8 & 22.2 & 21.5\nl 
\hline
\end{tabular}
\tablecomments{Units of right ascension (J2000) are hours, minutes, and 
seconds, and units of declination (J2000) are degrees, arcminutes, and 
arcseconds.}
\end{center}
\end{table}

\clearpage

\begin{deluxetable}{lccccc}
\tablecaption{\label{tab2} $RIZ$ member candidates in the $\sigma$\,Orionis 
cluster}
\tablewidth{0pt}
\tablehead{ 
\colhead{ } & \colhead{ }     & \colhead{ }  &
\colhead{ } & \colhead{R.A.}  & \colhead{Decl.}
\nl
\colhead{ID} & \colhead{IAU} & \colhead{$I$}  &
\colhead{$R$$-$$I$} & \colhead{(J2000)}  & \colhead{(J2000)} }
\startdata
SOri\,1 &SOri\,J053911.7--022741 & 15.08$\pm$0.04   &1.70$\pm$0.07   
&5 39 11.7   &--02 27 41\nl
SOri\,2 &SOri\,J053926.2--022838 & 15.12$\pm$0.04   &1.80$\pm$0.07   
&5 39 26.2   &--02 28 38\nl
SOri\,3 &SOri\,J053920.8--023035 & 15.16$\pm$0.04   &2.15$\pm$0.07   
&5 39 20.8   &--02 30 35\nl
SOri\,4 &SOri\,J053939.2--023227 & 15.23$\pm$0.04   &2.16$\pm$0.07   
&5 39 39.2   &--02 32 27\nl
SOri\,5 &SOri\,J053920.1--023826 & 15.40$\pm$0.05   &1.86$\pm$0.07   
&5 39 20.1   &--02 38 26\nl
SOri\,6 &SOri\,J053847.5--023038 & 15.53$\pm$0.04   &2.00$\pm$0.07   
&5 38 47.5   &--02 30 38\nl
SOri\,7 &SOri\,J053908.1--023230 & 15.63$\pm$0.04   &2.07$\pm$0.07   
&5 39 08.1   &--02 32 30\nl
SOri\,8 &SOri\,J053907.9--022848 & 15.74$\pm$0.04   &1.87$\pm$0.07   
&5 39 07.9   &--02 28 48\nl
SOri\,9 &SOri\,J053817.1--022228 & 15.81$\pm$0.04   &2.06$\pm$0.07   
&5 38 17.1   &--02 22 28\nl
SOri\,10&SOri\,J053944.4--022445 & 16.08$\pm$0.04   &1.97$\pm$0.07   
&5 39 44.4   &--02 24 45\nl
SOri\,11&SOri\,J053944.2--023305 & 16.15$\pm$0.04   &2.21$\pm$0.07   
&5 39 44.2   &--02 33 05\nl
SOri\,12&SOri\,J053757.4--023845 & 16.28$\pm$0.07   &1.94$\pm$0.09   
&5 37 57.4   &--02 38 45\nl
SOri\,13&SOri\,J053813.1--022410 & 16.35$\pm$0.04   &2.05$\pm$0.07   
&5 38 13.1   &--02 24 10\nl
SOri\,14&SOri\,J053937.5--024431 & 16.35$\pm$0.05   &2.23$\pm$0.08   
&5 39 37.5   &--02 44 31\nl
SOri\,15&SOri\,J053848.0--022854 & 16.55$\pm$0.04   &2.05$\pm$0.08   
&5 38 48.0   &--02 28 54\nl
SOri\,16&SOri\,J053915.0--024048 & 16.60$\pm$0.05   &2.15$\pm$0.08   
&5 39 15.0   &--02 40 48\nl
SOri\,17&SOri\,J053904.4--023835 & 16.80$\pm$0.05   &2.02$\pm$0.08   
&5 39 04.4   &--02 38 35\nl
SOri\,18&SOri\,J053825.6--023122 & 16.81$\pm$0.04   &2.11$\pm$0.08   
&5 38 25.6   &--02 31 22\nl
SOri\,19&SOri\,J053721.0--022543 & 16.81$\pm$0.04   &2.12$\pm$0.07   
&5 37 21.0   &--02 25 43\nl
SOri\,20&SOri\,J053907.4--022908 & 16.88$\pm$0.04   &2.12$\pm$0.08   
&5 39 07.4   &--02 29 08\nl
SOri\,21&SOri\,J053934.2--023847 & 16.90$\pm$0.05   &2.17$\pm$0.09   
&5 39 34.2   &--02 38 47\nl
SOri\,22&SOri\,J053835.2--022524 & 16.98$\pm$0.04   &2.24$\pm$0.08   
&5 38 35.2   &--02 25 24\nl
SOri\,23&SOri\,J053751.0--022610 & 17.00$\pm$0.04   &2.23$\pm$0.07   
&5 37 51.0   &--02 26 10\nl
SOri\,24&SOri\,J053755.6--022434 & 17.00$\pm$0.04   &2.15$\pm$0.07   
&5 37 55.6   &--02 24 34\nl
SOri\,25&SOri\,J053908.8--023958 & 17.04$\pm$0.05   &2.29$\pm$0.11   
&5 39 08.8   &--02 39 58\nl
SOri\,26&SOri\,J053916.6--023827 & 17.05$\pm$0.05   &2.04$\pm$0.09   
&5 39 16.6   &--02 38 27\nl
SOri\,27&SOri\,J053817.3--024024 & 17.07$\pm$0.05   &2.13$\pm$0.07   
&5 38 17.3   &--02 40 24\nl
SOri\,28&SOri\,J053923.1--024656 & 17.11$\pm$0.05   &2.15$\pm$0.09   
&5 39 23.1   &--02 46 56\nl
SOri\,29&SOri\,J053829.5--022517 & 17.18$\pm$0.04   &2.03$\pm$0.08   
&5 38 29.5   &--02 25 17\nl
SOri\,30&SOri\,J053913.0--023751 & 17.25$\pm$0.05   &1.90$\pm$0.09   
&5 39 13.0   &--02 37 51\nl
SOri\,31&SOri\,J053820.8--024613 & 17.31$\pm$0.05   &2.15$\pm$0.07   
&5 38 20.8   &--02 46 13\nl
SOri\,32&SOri\,J053943.5--024731 & 17.36$\pm$0.05   &2.04$\pm$0.09   
&5 39 43.5   &--02 47 31\nl
SOri\,33&SOri\,J053657.9--023522 & 17.38$\pm$0.04   &2.28$\pm$0.07   
&5 36 57.9   &--02 35 22\nl
SOri\,34&SOri\,J053707.1--023246 & 17.46$\pm$0.04   &2.01$\pm$0.07   
&5 37 07.1   &--02 32 46\nl
SOri\,35&SOri\,J053755.5--023308 & 17.50$\pm$0.04   &2.36$\pm$0.07   
&5 37 55.5   &--02 33 08\nl
SOri\,36&SOri\,J053926.8--023656 & 17.67$\pm$0.06   &2.18$\pm$0.14   
&5 39 26.8   &--02 36 56\nl
SOri\,37&SOri\,J053707.5--022717 & 17.70$\pm$0.04   &2.31$\pm$0.07   
&5 37 07.5   &--02 27 17\nl
SOri\,38&SOri\,J053915.1--022152 & 17.70$\pm$0.05   &2.13$\pm$0.10   
&5 39 15.1   &--02 21 52\nl
SOri\,39&SOri\,J053832.4--022958 & 17.82$\pm$0.05   &2.34$\pm$0.11   
&5 38 32.4   &--02 29 58\nl
SOri\,40&SOri\,J053736.4--024157 & 17.93$\pm$0.05   &2.34$\pm$0.07   
&5 37 36.4   &--02 41 57\nl
SOri\,41&SOri\,J053938.4--023116 & 18.44$\pm$0.06   &2.43$\pm$0.18   
&5 39 38.4   &--02 31 16\nl
SOri\,42&SOri\,J053923.3--024057 & 19.01$\pm$0.09   &2.47$\pm$0.41   
&5 39 23.3   &--02 40 57\nl
SOri\,43&SOri\,J053813.8--023504 & 19.05$\pm$0.04   &2.38$\pm$0.08   
&5 38 13.8   &--02 35 04\nl
SOri\,44&SOri\,J053807.0--024321 & 19.39$\pm$0.06   &2.31$\pm$0.15   
&5 38 07.0   &--02 43 21\nl
SOri\,45&SOri\,J053825.5--024836 & 19.59$\pm$0.06   &2.88$\pm$0.18   
&5 38 25.5   &--02 48 36\nl
SOri\,46&SOri\,J053651.7--023254 & 19.82$\pm$0.04   &2.61$\pm$0.11   
&5 36 51.7   &--02 32 54\nl
\enddata
\tablecomments{Units of right ascension (J2000) are hours, minutes, 
and seconds, and units of declination (J2000) are degrees, arcminutes, 
and arcseconds. Coordinates are accurate to $\pm$1\arcsec.}
\end{deluxetable}

\clearpage

\begin{deluxetable}{lcccccc}
\tablecaption{\label{tab3} $IZ$ ($I \ge 20$) member candidates in 
the $\sigma$\,Orionis cluster}
\tablewidth{0pt}
\tablehead{ 
\colhead{ } & \colhead{ }     & \colhead{ }  &  \colhead{ }  &
\colhead{ } & \colhead{R.A.}  & \colhead{Decl.}
\nl
\colhead{ID} & \colhead{IAU} & \colhead{$I$}  & \colhead{$I - Z$} &
\colhead{$R - I$} & \colhead{(J2000)}  & \colhead{(J2000)} }
\startdata
SOri\,47 & SOri\,J053814.5--024016 & 20.53$\pm$0.05 & 1.00$\pm$0.06 
& 2.4$\pm$0.3 & 5 38 14.5 & --02 40 16\nl
SOri\,48 & SOri\,J053739.0--023950 & 20.70$\pm$0.05 & 0.93$\pm$0.08 
& 2.2$\pm$0.2 & 5 37 39.0 & --02 39 50\nl
SOri\,49 & SOri\,J053822.9--023755 & 20.83$\pm$0.07 & 1.01$\pm$0.08 
& $\ge 1.5$   & 5 38 22.9 & --02 37 55\nl
\enddata
\tablecomments{Units of right ascension (J2000) are hours, minutes, 
and seconds, and units of declination (J2000) are degrees, arcminutes, 
and arcseconds. Coordinates are accurate to $\pm$1\arcsec. The $Z$ 
magnitudes have been calibrated as explained in Zapatero Osorio et al. 
(\cite{osorio98}). Finder charts are available upon request to the authors.}
\end{deluxetable}

\clearpage

\begin{deluxetable}{lccccccccccc}
\footnotesize
\tablecaption{\label{tab4} Spectroscopic data and fundamental parameters}
\tablewidth{0pt}
\tablehead{
\colhead{ID}  & 
\colhead{$I$}   & 
\colhead{SpT}   & 
\colhead{Na~{\sc i}}   & 
\colhead{H$\alpha$} & 
\colhead{$\log L/L_{\odot}$}   & 
\colhead{$M/M_{\odot}$} & 
\nl
\colhead{}      & 
\colhead{}      & 
\colhead{}      & 
\colhead{(\AA)} & 
\colhead{(\AA)} &        
\colhead{($\pm 0.15$)}      & 
\colhead{(1--10\,Myr)}      & 
                            }
\startdata
SOri\,12 & 16.28 & M6.0 &2.5      &6.5 &--1.45&0.070--0.200\nl
SOri\,17 & 16.80 & M6.0 &$\le 1.0$&5.5 &--1.65&0.055--0.150\nl
SOri\,29 & 17.18 & M6.0 &2.2      &28.0&--1.81&0.045--0.110\nl
SOri\,25 & 17.04 & M6.5 &$\le 1.0$&45.0&--1.72&0.050--0.130\nl
SOri\,39 & 17.82 & M6.5 &$\le 1.0$&5.1 &--2.02&0.040--0.080\nl
SOri\,27 & 17.07 & M7.0 &$\le 1.0$&6.1 &--1.66&0.055--0.150\nl
SOri\,40 & 17.93 & M7.0 &$\le 1.0$&30.0&--2.00&0.040--0.080\nl
SOri\,44$^{\ast}$ & 19.39 & M6.5 & 5.0 & $\le$5 & -- & --  \nl
SOri\,45 & 19.59 & M8.5 &$\le 1.0$&60.0&--2.40&0.020--0.040\nl
\enddata
\tablecomments{Uncertainties in spectral types are half a subclass; in 
EWs they are $\pm$1\,\AA.\\
$^{\ast}$ \ This object does not show spectroscopic features distinctive 
of the youth of the $\sigma$\,Orionis cluster, and therefore it has not been 
considered for the determination of luminosity and mass. }
\end{deluxetable}

%%%
%%%  FIGURES
%%%

\clearpage
\plotone{bejar1.ps}
\clearpage
\plotone{bejar2.eps}
\clearpage
%\plotone{bejar3a.eps}
%\clearpage
%\plotone{bejar3b.eps}
%\clearpage
\plotone{bejar4.eps}
\clearpage
\plotone{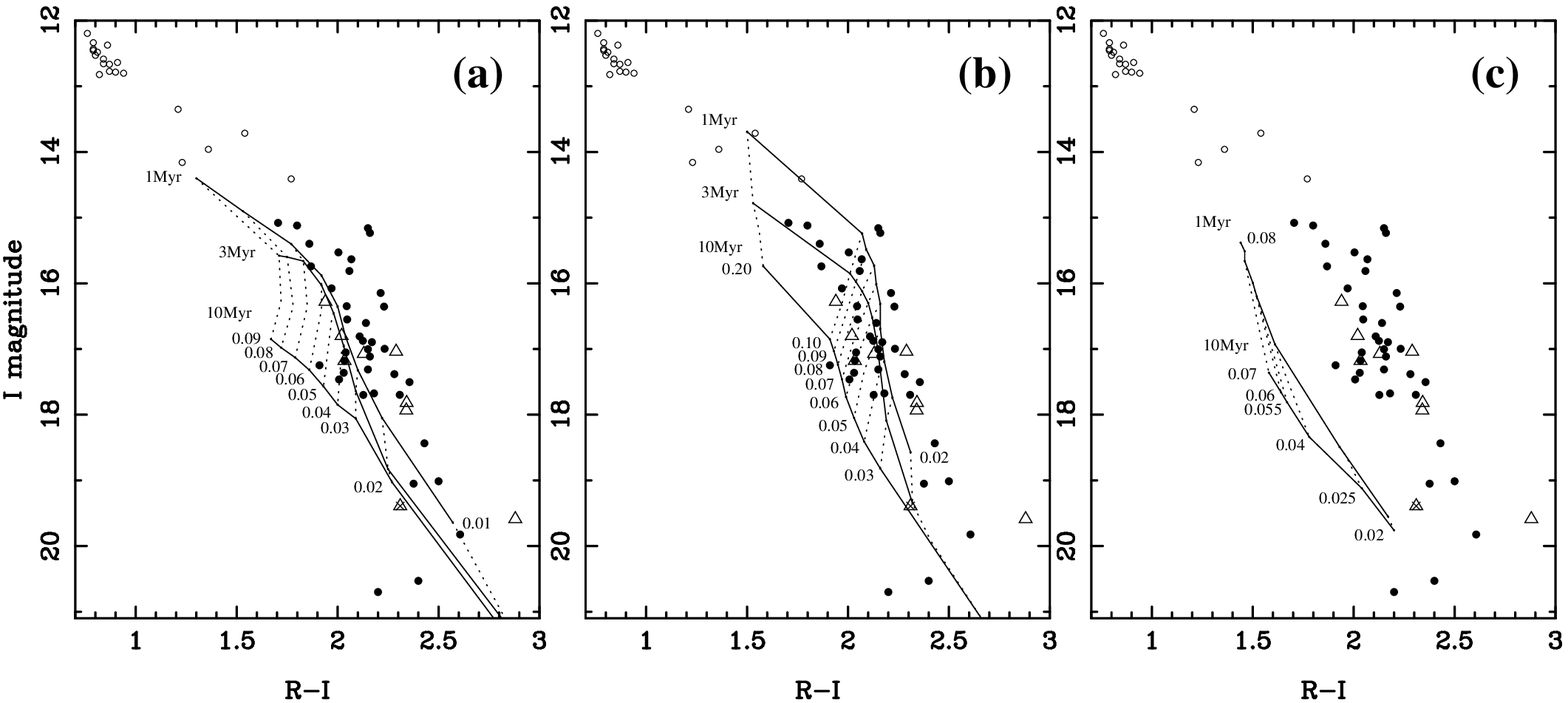}
\clearpage
\plotone{bejar6.eps}
\clearpage
\plotone{bejar7.eps}
\clearpage
\plotone{bejar8.eps}
\clearpage
\plotone{bejar9.eps}

\end{document}